\documentstyle[twoside,fleqn,espcrc2]{article}

\def\eq#1{Eq.~(\ref{#1})}

\def\beq{\begin{equation}}
\def\eeq{\end{equation}}
\def\beqa{\begin{eqnarray}}
\def\eeqa{\end{eqnarray}}

\newcommand{\as}{\alpha_s}

\newcommand{\AmS}{{\protect\the\textfont2
  A\kern-.1667em\lower.5ex\hbox{M}\kern-.125emS}}
\catcode`@=11 
\def\gsim{\mathrel{\rlap{\lower4pt\hbox{\hskip1pt$\sim$}}
    \raise1pt\hbox{$>$}}}         
\def\lsim{\mathrel{\mathpalette\@versim<}}
\def\@versim#1#2{\lower0.2ex\vbox{\baselineskip\z@skip\lineskip\z@skip
       \lineskiplimit\z@\ialign{$\m@th#1\hfil##$\crcr#2\crcr\sim\crcr}}}
\catcode`@=12 
\def\epm#1#2{\hbox{${\lower1pt\hbox{$\scriptstyle +~#1$}}
\atop {\raise1pt\hbox{$\scriptstyle -~#2$}}$}}
\def\blan{\Big\langle}
\def\bran{\Big\rangle}

\title{A determination of $\alpha_s$ from scaling violations with 
truncated moments}

\author{S. Forte\address{INFN, Sezione di Roma Tre,\\
Via della Vasca Navale 84, I-00146 Rome, Italy},
J.I. Latorre\address{Departament d'Estructura i Costituents de 
la Mat\`eria, \\ Universitat de Barcelona, Diagonal 647, E-08028 
Barcelona, Spain},
L. Magnea\address{Dipartimento di Fisica Teorica,
Universit\`a di Torino and INFN, Sezione di Torino,\\
Via P. Giuria 1, I-10125 Torino, Italy} and
A. Piccione\address{Dipartimento di Fisica, Universit\`a di Genova 
and INFN, Sezione di Genova,\\ Via Dodecaneso 33, I-16146 Genova, Italy}}

\begin{document}

\begin{abstract}

We describe a determination of the strong coupling $\alpha_s (M_Z)$
from scaling violations of the nonsinglet DIS structure function,
which is based on two novel techniques aimed at controlling and
minimizing the theoretical error: a neural network parametrization of
BCDMS and NMC data, and QCD evolution by means of truncated Mellin
moments.

\end{abstract}

\maketitle

\section{Introduction}
\label{intr}

Scaling violations of DIS structure functions play a key role in our
understanding of perturbative QCD, and their measurement is one of the
most natural and potentially cleanest methods available for the
determination of the strong coupling~\cite{beth}. There are, however,
practical and conceptual difficulties that must be overcome to achieve
with this method a precise determination of $\alpha_s$ and its
associated statistical and theoretical errors.

The theoretical prediction for scaling violations is given by
Altarelli--Parisi evolution equations~\cite{AP}. A clean, analytic
solution at NLO in the coupling is available in terms of Mellin
moments of structure functions, which diagonalize the evolution
operator. The difficulty in the implementation of this solution is the
fact that moments are not directly measurable, since they are weighted
integrals of the structure function over the interval $0 < x < 1$, and
$x \to 0$ implies $\sqrt{s} \to \infty$. The uncertainty on the value
of any given moment is thus unknown, and in principle infinite: it is
necessary to rely upon extrapolations. An alternative, practical, if
less elegant method of solution is to work numerically, directly with
the $x$--space integro--differential equation. This involves, in
principle, no extrapolation, since AP evolution is directional in $x$
space, however it can be numerically challenging. Furthermore, in
practice, this method usually requires the introduction of a parton
parametrization, and this raises new issues: in fact, the choice of a
parametrization is a source of theoretical bias, very difficult to
assess, and errors on physical observables associated with the
parametrization are notoriously difficult to evaluate~\cite{qcd}.

Our goal here is to present a data--driven determination of 
$\alpha_s$~\cite{us}, which tackles the problems outlined above. We 
use BCDMS~\cite{BCDMS} and NMC~\cite{NMC} data for the nonsinglet 
DIS structure function $F_2$, and we strive to minimize all sources 
of theoretical bias and uncertainty, while accurately assessing the 
statistical effects of experimental errors and correlations.

To this end, we employ two theoretical tools which have recently been
developed. First, we solve AP equations by means of truncated Mellin
moments~\cite{us1}; next, we make use of a bias--free parametrization
of $F_2$, constructed by means of neural networks~\cite{us2}.

Mellin moments over a truncated interval ($x_0 < x < 1$) are
observable quantities; as shown in Ref.~\cite{us1}, they obey a simple
evolution equation, which can be approximated with arbitrary precision
by a matrix equation admitting a simple analytic solution. In
principle, since one is manipulating observable quantities, a
parametrization of the structure function is not needed.  In practice,
however, data coverage and precision are not sufficient, and to
extract from the data the maximum amount of information (for example,
in order to combine errors on neighboring data points in the best way)
it is necessary to impose the constraint of continuity on the
structure function, {\it i.e.} to introduce a parametrization. To do so
without introducing a theoretical bias, and keeping full control on
the propagation of experimental errors and correlations, we make use
of the results of Ref.~\cite{us2}, where neural networks were employed
in conjunction with Monte--Carlo techniques to construct a faithful
representation of the probability distribution in the space of
structure functions.

\section{Evolution with truncated moments}
\label{trun}

Truncated Mellin moments of a parton distribution $q(x, t)$, with
$t = \log \mu_f^2$, are defined by
\beq
q_n (x_0, t) \equiv \int_{x_0}^1 d x ~x^{n - 1} q(x, t)~.
\label{trumo}
\eeq
They satisfy the evolution equation
\beq
\frac{d q_n}{d t} = \frac{\alpha_s}{2 \pi} \int_{x_0}^1 d y ~y^{n - 1} 
q(y, t) ~G_n \left(\frac{x_0}{y}, \alpha_s \right),
\label{truncalpar}
\eeq
where
\beq
G_n (x, \alpha_s) = \int_x^1 d z z^{n - 1} P(z, \alpha_s)
\label{kern}
\eeq
is the truncated moment of the appropriate AP kernel $P(z, \alpha_s)$.

As $x_0 \to 0$, $G_n$ becomes the anomalous dimension $\gamma_n$, and
different moments evolve independently; for $x_0 \neq 0$, evolution
couples the $n$-th truncated moment $q_n$ with all other $q_k$, with
$k > n$, as easily seen by Taylor expanding $G_n(x_0/y)$ around $y =
1$.  This Taylor expansion converges in the interval $x_0 < y
\leq 1$, since $G_n$ only has integrable singularities due to $+$
distributions at $y = x_0$. Truncating the expansion at the $M$-th
term yields the linear system
\beq
\frac{d q_n}{d t} =
\frac{\alpha_s}{2 \pi} \sum_{p=0}^{M} ~c_{p,n}^{(M)}(x_0,
\alpha_s) ~q_{n + p}(x_0, t)~,
\label{finsyst}
\eeq
with coefficients that can be computed analytically to the order to
which the splitting functions are known.

Methods for the solution of \eq{finsyst} and their accuracy have been
studied in Refs.~\cite{us1,us3,deer}. One may first of all observe
that the matrix of anomalous dimensions governing the evolution of
truncated moments is upper triangular. As a consequence, it can be
diagonalized analytically by means of a simple recursion relation.  A
second observation is that moments with significantly different
indices are weakly coupled for small $x_0$. This is what justifies the
truncation of the expansion of the {\it r.h.s.} of \eq{truncalpar} at
finite $M$.

The convergence of the series of approximations for increasing $M$ has
been studied systematically. The convergence of the approximation as a
function of $M$ is good (leading to a few percent error for $M
{\scriptstyle \lsim} 20$), except for lowest nonsingular moments
(sensitive to singularities at $y = x_0$). An improved version of the
method~\cite{deer} deals with this problem, so that in fact for all
finite moments $M {\scriptstyle \lsim} 12$ is sufficient to achieve an
accuracy on evolution at the percent level.

The method has also been extended to singlet and gluon
distributions~\cite{us3}, with minor technical complications; a NLO
analytic solution is available in all cases and can be efficiently
implemented numerically (for details, see~\cite{antes}). It is worth
emphasizing that, being based on Mellin moments, the method is also
well suited to include the effects of threshold logarithms, which may
in fact play a non--negligible role in the determination of
$\alpha_s$.

\section{Neural network parametrization of $F_2$}
\label{neur}

The standard procedure for fitting structure functions, as well as
parton distributions, is to choose a simple functional form with
enough free parameters, and then fix the value of the parameters by
minimizing a suitable $\chi^2$. This procedure leads to well known
difficulties in determining errors to be associated with generic
observables depending on the fitted functions. First of all, the
computation of errors and correlations of the fit parameters requires
at least a fully correlated analysis of experimental errors, which is
not always available, and may be very difficult when data from
different experiments are involved. Furthermore, standard error
propagation methods are not generally applicable: many observables are
strongly nonlinear functionals of the fit parameters, and they often
depend on nonlocal features of the fitted functions. Finally, as
mentioned above, the theoretical bias due to choice of parametrization
is difficult to assess, and its effects can be large if data are not
precise, as it happens for example in the case of polarized
distributions~\cite{pdfrev}.

To solve these problems, what is needed~\cite{pdfer} is a reliable
representation of the probability measure ${\cal P}(F_2)$ in the space 
of structure functions $F_2(x, Q^2)$. Then, for any observable 
functional ${\cal G}(F_2)$, one could compute
\beq
\blan {\cal G} \left[ F_2 \right] \bran = \int \!
{\cal D} F_2 \, ~{\cal G} \left[ F_2(x,Q^2) \right] \, {\cal P}(F_2),
\label{funave}
\eeq
and similarly for higher moments. Just such a representation was 
constructed, using a combination of Monte--Carlo techniques and 
neural nerworks, in Ref.~\cite{us2}.

Neural networks are a class of algorithms providing robust, universal
and unbiased approximants to incomplete or noisy data. As such, they
are ideally suited to handle problems like the ones discussed above,
which center on the need to reconstruct a continuous function from a
discrete set of data with errors.

Space prevents us from discussing here in any detail the technology of
neural networks, their properties and applications (for an
introduction, see for example Ref.~\cite{netrev}). For the record, the
neural architecture implemented in Ref.~\cite{us2} is that of a
multilayer feed--forward network or ``perceptron''. This means that
network nodes (neurons) are arranged in an ordered sequence of layers,
and each neuron receives input from the neurons in the preceding
layer, while feeding output to those in the successive layer.  The
network learns to interpolate the chosen set of data by the method of
supervised training by back--propagation. In practice, the network
attempts matching data to output, then proceeds to vary the parameters
characterizing neuron activation (weights and thresholds), along the
steepest descent contour, searching for the minimum of the chosen
error function.

The only assumption made by the network algorithm is that the data can
be interpolated by a smooth function. All the parameters
characterizing the network, such as size (number of neurons),
architecture (structure of layers), length of the learning cycle, can
be decided based on purely statistical criteria. Specifically, no
assumption is made on the functional form of the interpolating
function.

The parametrization of the nonsinglet structure function
$F_2^{(NS)}(x, Q^2)$ constructed in Ref.~\cite{us2} is based on a
total of $552$ data points collected by the NMC and BCDMS
collaboration. The method used is a combination of Monte--Carlo
techniques with neural network technology, and consists of
two steps. The first step is the Monte--Carlo generation of an
ensemble of $N_{{\rm rep}}$ pseudo--data sets, mimicking the real
data, with the correct multivariate distribution given by experimental
errors, fully correlated. Let $i \equiv \{x, Q^2\}$ be a point where
$F_2^{(NS)}$ has been measured, obtaining the result $F_i^{(exp)}$,
with statistical error $\sigma_{i,s}$, normalization error
$\sigma_{i,N}$ and percentage systematic errors $f_{i,a}$. Then one
generates the $N_{{\rm rep}}$ pseudo--data at point $i$ according to
\begin{eqnarray}
F_i^{(art)\,(k)} & = & \left(1 + ~r_{i,N}^{(k)} ~\sigma_{i,N}\right) 
\Bigg[ F_i^{(exp)} \\
& + & \frac{ \sum_a r_{i,a}^{(k)} f_{i,a}}{100} ~F_i^{(exp)} +
r_{i,s}^{(k)} ~\sigma_{i,s} \Bigg]~, \nonumber
\label{gennmc}
\end{eqnarray}
where $k = 1, \ldots, N_{{\rm rep}}$, and $r_{i,a}^{(k)}$,
$r_{i,N}^{(k)}$ and $r_{i,s}^{(k)}$ are univariate gaussian random
numbers, which for systematic errors are grouped in classes according
to experimental correlations.

The second step is the training of $N_{{\rm rep}}$ neural networks, each 
one using one pseudo--data set. At the end of training, the parameters
of each network are optimized for the interpolation of the corresponding 
data set; the output of the process is a set of $N_{{\rm rep}}$ continuous 
functions $F_{{\rm net}}^{(k)} (x, Q^2)$ representing a faithful sample
of the probability distribution ${\cal P}(F_2)$.

Finally, one may evaluate averages, errors and correlations of observables 
using the $N_{{\rm rep}}$ networks as a Monte--Carlo representation of 
${\cal P}(F_2)$. In this context, $\alpha_s$ can be treated like any 
functional of the structure function.

\section{Fit architecture and parameters}
\label{fita}

Having at our disposal the ensemble of structure functions needed for
the analysis, there are still a number of parameters that may be
chosen in the fitting procedure to determine $\alpha_s$, to maximize
the accuracy of the results. First of all, we must choose the
truncation point $x_0$ entering the definition of truncated moments,
\eq{trumo}, as well as the fitting range in $Q^2$. The criteria
for these choices are simple: we must take maximal advantage of data
coverage, in order to minimize statistical errors on individual
moments; we must impose a lower cut on $Q^2$ to keep power correction
under control; we must keep $x_0$ as small as possible, compatibly
with data coverage, to ensure a fast convergence of our approximate
evolution equation. We also have a choice regarding the number of
intermediate scales to be used as evolution targets: in principle we
could apply the evolution equation between any two scales in the
fitting range, however introducing too many intermediate scales would
clearly not add new information, and it would lead to large
correlations between neighboring moments. Based on these criteria and
extensive testing, we choose $x_0 = 0.03$ for the truncation point,
$20 ~{\rm Gev}^2 < Q^2 < 70 ~{\rm Gev}^2$ for the fitting range, and
$n_{{\rm sc}} = 3$ for the number of scales, which are taken to be
equally spaced on a logarithmic scale.

A second set of choices to be made concerns our approximate evolution
equation. We work with a NLO evolution equation with matching at heavy
quark thresholds according to the Marciano prescription~\cite{marc};
we must then decide how many truncated moments should be included in
the evolution to achieve a satisfactory accuracy while preserving
numerical stability. An optimal choice is to include truncated moments
with $1 \leq n \leq 11$, {\it i.e.} to set $M = 11$ in
\eq{truncalpar}.  Since we are including the lowest
nonsingular moment, we must use the improved method of solution
described in Ref.~\cite{deer}; the auxiliary parameter introduced
there is set to $N = 6$. We performed several tests on our approximate
solution with these parameters, and found that the accuracy achieved
on evolution in the relevant range is at the level of $0.1 \%$.

A final important issue is the number of moments that should be
treated as fit parameters together with the strong coupling. In fact,
the values of truncated moments at the target scales of perturbative
evolution depend both on the coupling and on the values of the same
moments at the initial scale. These initial values can be either fixed
at the central experimental value, or fitted. It turns out that to
achieve a satisfactory precision, given the statistical errors on
truncated moments, the number of fitted moments must be $n_{{\rm fit}}
> 3$. On the other hand, fitting a large number of moments, especially
successive ones, endangers the numerical stability of the fit due to
the very large correlations between neighboring moments. This forces
$n_{{\rm fit}} < 6$. After testing the possible combinations, we find
that the choice that minimizes errors while maintaining numerical
stability is to fit moments $n = 2,4,5,6,8$.

We emphasize that all fit parameters have been varied within their
respective windows of stability with negligible effects on the
results.

\section{Result and errors}
\label{twol}

As shown in Ref.~\cite{us2}, an ensemble of $N_{{\rm rep}} = 1000$
networks is sufficient to reproduce correctly all experimental errors
and correlations, without introducing any bias. With such an ensemble,
and the fit architecture outlined above, our result with statistical
errors is
\beq
\as(M_Z) = 0.124 ~\epm{0.004}{0.007} ~\hbox{(stat.)}~.
\label{alstat}
\eeq
Theoretical uncertainties remain to be estimated. A first source is
the presence of power correction, which are not accounted for by a
perturbative treatment. They can be of a kinematical nature (target
mass corrections), or dynamical (higher twist corrections), or due to
elastic contributions at $x = 1$. As discussed in Ref.~\cite{us}, all
are found to be negligible ($ < 1\%$) thanks to our choice of $Q^2$
range. A second source of theoretical error is given by the
uncalculated NNLO and higher perturbative contributions to the
evolution equation.  These can be estimated by varying the
renormalization scale according to $\mu^2_{ren} = k_{ren} Q^2$ (note
that there is no factorization scale dependence in DIS scheme). We
have tested the range $0.3 < k_{ren} < 4$, and found that the ensuing
uncertainty is not negligible, indicating sizeable NNLO corrections:
$\sigma_{{\rm ren}} = \epm{0.003}{0.004}$.  It is conceivable, and it
can be tested, that our method might be affected by an enhancement of
threshold logarithm effects, since the fitting procedure involves
relatively high Mellin moments. The inclusion of such logarithms in a
resummed fit should be natural with our formalism, since resummation
is performed in Mellin space, where our evolution takes place.  A final
possible source of theoretical error is the location of heavy quark
thresholds. This is estimated by varying the threshold position as
$Q_{th}^2 = k_{th} M_q^2$, with $0.3 < k_{th} < 4$. The effect is expected
to be nearly negligible, since only the $b$ threshold is included in our
$Q^2$ range, and only for some $k_{th}$. We find in fact that
$\sigma_{{\rm th}} = \epm{0.000}{0.002}$.

Summarizing, our final result for the strong coupling reads
\beq
\as(M_Z) = 0.124 ~\epm{0.004}{0.007}~\hbox{(exp.)} ~\epm{0.003}{0.004} 
{} ~\hbox{(th.)}. 
\label{alfin2} 
\eeq
We observe that the error is dominated by statistical uncertainties,
consistently with our expectations. The central value turns out to be
somewhat on the high side of the current world average, though well
within errors. We note that this is consistent with the possibility
that threshold logarithms might affect our determination more than
others, since it is known that their leading effect is to replace the
argument of the coupling, changing $Q^2$ to $Q^2/n$ for the $n$--th
moment, thus leading to a larger value for the effective coupling.

Resummation of soft gluon effects will probably lead to a further
reduction of the theoretical error, which is dominated by unknown
higher order corrections. On the experimental side, the result could
be significantly improved either with better statistical accuracy
(particularly of deuteron data), or by extending the range in $Q^2$,
which might be achieved by the planned EIC facility~\cite{eic}.

\vspace{3mm}

\noindent {\large {\bf Acknowledgements}} 

\vspace{1mm}

\noindent We thank L. Garrido, co--author of Ref.~\cite{us2}, for many
useful discussions.  Work supported in part by EU TMR contract
FMRX--CT98--0194 (DG 12 -- MIHT) and by Spanish and Catalan grants
AEN99--0766, AEN99--0483, 1999SGR--00097.

\end{document}